\documentclass[11pt,twoside]{article}
\usepackage{asp2006}

\begin{document}
\title{Searching for Planetary Transits in Star Clusters}
\author{David T.F Weldrake}
\affil{Max Planck Institut f\"ur Astronomie, K\"onigstuhl 17, Heidelberg, D-69117 Germany. }

\begin{abstract}
Star clusters provide an excellent opportunity to study the role of environment on determining the frequencies of short period planets. They provide a large sample of stars which can be imaged simultaneously, with a common distance, age and pre-determined physical parameters. This allows the search to be tailor-made for each specific cluster. Several groups are attempting to detect transiting planets in open clusters. Three previous surveys have also targeted the two brightest globular clusters. No cluster survey has yet detected a planet. This contribution presents a brief overview of the field, highlighting the pros and cons of performing such a search, and presents the expected and current results, with implications for planetary frequencies in regions of high stellar density and low metallicity.
\end{abstract}

\section{Introduction - Why Study Star Clusters?}   
Sampling star clusters (both open and globular) to detect transiting planets provides unique opportunities to study the frequencies and populations of short period giant planets in differing regions of the Galaxy. They allow investigation into the effects of stellar density and metallicity on planet occurrence, particularly in regions too distant to be sampled by ground-based radial velocity surveys. 

Star clusters provide a large sample of main sequence dwarf stars which can be simultaneously imaged with wide fields of view over long periods. This results in sufficient photometric precision for the statistical detection of transiting giant planets with only moderate ground-based instruments. The sampled stars will have a common age, distance and origin. By fitting the stellar population with theoretical isochrones, this allows investigation into the stellar physical properties. This is of great importance to tailor a transit search for any particular cluster, maximizing transit detection and permitting a detailed understanding of the statistical significance of any null result.

Particular scientific objectives are to place upper limits on the timescales of planetary formation and subsequent migration, to study the dependence of planetary mass and radius with stellar mass (if found in large numbers), and investigate the trend of planetary frequency with stellar metallicity. Comparisons can also be made between observed planetary frequencies and the simulated effects of stellar encounters on planetary formation in regions of high stellar density. The clusters themselves must be chosen carefully, with those containing many thousands of low-mass members in bright and nearby associations having statistically the best chance of success. Such clusters are relatively rare.

\section{Transits in Open Clusters - Young Planets and `Hot Earths'}   
Several groups have attempted the search for transiting planets in selected open clusters. Examples are NGC 6791 \citep{B2003}, NGC 6819 \citep{S2003}, NGC 7789 \citep{B2005}, NGC 2260 and NGC 6208 \citep{vB2005}, NGC 7086 \citep{R2006}, NGC 2158 \citep{M2006} and NGC 1245 \citep{Bu2006}. Several transit candidates have been identified during the course of these studies, but due to their faint nature, none have yet been confirmed as transiting planets.

Per cluster, the chance of detecting a planetary transit is low, based on the relatively small number of stars available for study. On the other hand, as open clusters are young objects, only the most massive stars will have evolved from the main sequence. The vast majority of the stellar population would hence be late-type dwarf stars suitable for transit detection. 

As there is tentative evidence that lower mass stars harbor on average lower mass planets (and perhaps also in greater numbers than high mass planets), open clusters can be used as targets to detect the first transiting planet approaching the Earth in radius. With careful planning of the observing strategy and choice of cluster, along with those tailored analysis methods determined via simulated data \citep[see][]{PG2005,AP2007}, cluster detection prospects can be greatly increased. 

\subsection{Surveys and Future Potential}  
Clearly the most effective way to maximize the chance of a detection is to observe as many clusters as possible. Several large-scale projects are underway, targeting the most promising bright, nearby and rich open clusters. Ongoing projects include UStAPS \citep{St2003}, EXPLORE-OC \citep{vB2005}, $\it{monitor}$ \citep{A2007}, PISCES \citep{M2002} and STEPSS \citep{B2004}. These projects continue to deliver promising transit candidates, cluster parameters and catalogs of variable stars. 

Depending on intrinsic planet occurrence frequencies, \citealt{PG2006} have shown via simulated data that dedicated ground-based transit surveys of the nearest, richest and brightest open clusters, with moderate instruments lasting $\ge$20 nights, have great potential to detect both transiting `Hot Neptune', and `Hot Earth' planets in association with low mass stars. The near future no doubt holds the first discovery of a young transiting planet in a metal-rich open cluster. Subsequent surveys will hence mirror a general trend in transit astronomy - the detection of ever smaller planets.

\section{Transits in Globular Clusters - Testing Environment and Occurrence Rates}
Globular Clusters are excellent targets for transit surveys. They provide thousands of upper main sequence stars ($\sim$solar mass) in a single field of view ($\sim$1 deg$^2$), and as such provide excellent statistics for planet detection, and a detailed understanding of any null result. Globulars are a very different environment to those regions surveyed for planets with radial velocities, or indeed in open clusters, having low metallicity and being highly crowded.

Two globulars, 47 Tucanae (47 Tuc) and $\omega$ Centauri ($\omega$ Cen), are bright and rich enough to be studied in detail with small to moderate ground-based instruments. Both have been surveyed for transits, . For 47 Tuc, the core was sampled with HST for 8.3 days by \citet{G2000} and 22,000 stars in the outer halo in a 33 night ground-based search by \citet{W2005}. When combined, the sheer numbers of stars from these two projects permits the detection of 24 planets in 47 Tuc. None were seen; a null result of high significance. As \citet{W2005} surveyed the uncrowded outer halo, the indication is that system metallicity is dominating over cluster dynamics in reducing the occurrence rate of short period giant planets in 47 Tuc. 

For $\omega$ Cen, \citet{W2007} sampled 31,000 main sequence stars in the cluster halo for 27 nights in a similar manner to 47 Tuc. A total of 2-5 planets were expected and again none were found, leading to the determination of upper limits to the occurrence rate of large-radius ($R=1.5R_{\rm{Jup}}$) short period planets in this cluster to 95$\%$ confidence. Again, metallicity was held responsible for the low frequency, as stellar density was dynamically speaking far less of an issue, but just how do stellar encounters affect planet survivability in these environments?

\subsection{The Globular Cluster Environment: Stellar Encounters and Planet Survivability}
Any hypothetical globular cluster planet would have to contend with conditions unlike anything in the solar neighborhood. For 47 Tuc, \citet{S1992} and \citet{DS2001} showed that a planet in the densest core region of 47 Tuc, sampled by \citet{G2000}, would survive disruption by stellar encounters for $\sim1 \times 10^{8}$ yrs at 1 AU semi-major axis. Short period planets, those to which transit surveys are sensitive, would survive for significantly longer. 

For the uncrowded cluster halo sampled by \citet{W2005}, \citet{B2001} showed that planets with 10 AU semi-major axis would still be intact, hence transiting planets at $\sim$0.04 AU would be remarkably unaffected by cluster dynamics. $\omega$ Cen has five times the mass of 47 Tuc, yet only 1/10$^{\rm{th}}$ the core density, constituting a far more planet-friendly environment (survivability at 10 AU $\ge$10$^{10}$ yrs), yet still no planets were found.

For both clusters, the conclusion was that if short period giant planets formed there in the first place, with the same frequency as in the solar neighborhood, then they would remain to be detectable in large numbers in transit surveys. It is well known that stellar metallicity plays an important r\^ole in determining the frequencies of giant planets in the solar neighborhood \citep{FV2005}. The results of the globular cluster transit surveys show that this trend is also present at far lower metalicities than those probed by radial velocity surveys. More distant metal rich globular clusters (eg: Terzan 5) should hence harbor detectable planets, but ground-based studies are unable to reach such distant and obscured clusters.

\section{Summary and Conclusions}
Identifying transiting planets in star clusters (both open and globular) presents many unique opportunities to study the formation, evolution, occurrence frequency and population of short period giant planets in regions unreachable by radial velocity surveys. They provide a large sample of dwarf stars with a common origin, distance and age, that can be simultaneously imaged with modest instruments from the ground.

Open clusters can reveal young and/or small planets, with multiple targets chosen to maximize potential detections. Some of the brightest, richest and most nearby open clusters have been subjected to transit searches. Despite no planets yet being confirmed, these surveys have provided promising candidates awaiting follow-up, as well as cluster parameters and variable stars. From extensive simulations, open clusters provide excellent opportunities to detect `Hot Neptune' planets, and the first transiting `Hot Earth' in the near future.

The two brightest globular clusters have also been surveyed, both from space and from the ground. Due to the large number of sampled stars, 26-29 planets in total should have been identified. None were found. Dynamical simulations have shown that stellar metallicity is very likely the cause for the paucity of short period giant planets, the metallicity dependence on planet formation is hence very likely operating at metalicities far lower than those probed in the solar neighborhood.
  
\acknowledgements 
DTFW would like to thank the conference organisers for the invitation to attend and present this review.

\end{document}